\documentclass[preprint]{revtex4}
\usepackage[dvips]{graphicx}
\usepackage{epsfig}
\begin{document}

\title{ Symmetries of Electrostatic Interaction between  DNA Molecules}

\author{Voislav Golo}
\email{golo@mech.math.msu.su }

\author{Efim Kats}
\email{kats@ill.fr}

\author{Yuriy Volkov}
\email{yu.volkov@gmail.com}

\affiliation{
       $^{*}$$^{\ddag}$Department of Mechancs and Mathematics \\
        Moscow University,  Moscow 119 899, Russia;\\
       $^{\dag}$Institute Landau, Moscow, Russia, and\\
        Institut Laue-Langevin,  Grenoble, France}
\date{July 4,  2007}

\begin{abstract}
We study a model for pair interaction $U$ of DNA molecules
generated by the discrete dipole moments of base-pairs and the
charges of phosphate groups, and find  noncommutative group of
eighth order ${\cal S}$ of symmetries that leave $U$ invariant. We
classify the minima  using group ${\cal S}$ and employ numerical
methods for finding them. The minima may correspond to several
cholesteric phases, as well as phases formed by cross-like
conformations of molecules at an angle close to $\rm{90}^{o}$ ,
"snowflake phase". The results depend on the effective charge $Q$
of the phosphate group which can be modified by the polycations or
the ions of metals. The snowflake phase could exist for $Q$ above
the threshold $Q_C$. Below $Q_C$ there could be several
cholesteric phases. Close to $Q_C$ the snowflake phase could
change into the cholesteric one at constant distance between
adjacent molecules.
\end{abstract}


\maketitle

\section{\label{sec:intro}Introduction.}

The DNA cholesterics have drawn considerable attention as unusual states of condensed matter,
\cite{robinson} - \cite{yevdokimov}. These liquid-crystalline phases are of interest from various
points of view, \cite{y1}. For one thing there is  a certain affinity, as packaging is concerned,
between them and viruses and chromosomes, for another they could be useful for constructing
nano-devices, for example sensors, \cite{y2}.

Cholesteric assemblies of the DNA have been observed in vivo and extensively studied in vitro,
\cite{yevdokimov}-\cite{lk1}. The experimental methods for obtaining molecules of the DNA in condensed
state  employ either solutions of polycations, or of polymers separating phases containing molecules
of the DNA.  Livolant et al, \cite{fl2004}, \cite{fl2006}, used polycations and found the spatial
characteristics of the location of DNA molecules, \cite{gross}. Yevdokimov et al, \cite{y1}, have used
the PEG pellets for the formation  of liquid crystalline phases of the DNA, and found the polymorphism
by observing a change in the sign  of cholesteric spiral.

By now it is generally accepted that the liquid crystalline phases of the DNA are essentially of the
electrostatic nature due to the helical distribution of charge, \cite{lk_model},  the conformational
properties of the DNA playing the crucial role, \cite{lk1}, \cite{lk2} . The concept of the charged
duplex has made for understanding the physics of the phases, and  enabled to explain the emergence of
the cholesteric structure and its relation to the ion strength of solutions,\cite{lk3} .

But the existence of different cholesteric phases of the DNA is
still a challenge to the current theory. The theoretical  study of
pair-interaction $U$ caused by charges of phosphate groups,
dipoles of base-pairs, and ions of ambient solution, can make for
understanding the phenomenon. Electric charges of molecules of the
DNA are of two kinds: those due to the phosphate groups and those
of the base-pairs, which are generally described by dipole
moments, \cite{sponer}.  Kim, \cite{kim}, used the helix symmetry
of the set of dipole moments, for describing a torque that may
deform the parallel conformation of two adjacent molecules and
result in a cholesteric phase of the system (see also
\cite{osipov}). The idea turned out to be quite useful for the
physics of the DNA. The main point about its implementation is
that the discrete structure of charges presents serious
difficulties for analytical treatment.

\section{\label{sec:main}Pair interaction of DNA molecules.}

In this paper we study the multistability of cholesteric phases, \cite{y2}, within the framework of
the approach indicated above, taking into account only the electrostatic forces between molecules of
the DNA in solution. We use the discrete approximation for the electrostatic interaction between pairs
of molecules of the DNA, \cite{kim}, use extensively computer simulation, and employ symmetry
arguments. Our model is essentially that of one dimensional lattice of charges and dipoles with an
elementary cell of size $3.4 \; \AA$ that mimics the spatial conformation of the charges of phosphate
groups and the dipoles of base-pairs.  We consider short segments of the DNA, approximately $500 \;
\AA$, that is of the size of persistence length, so that to a good approximation one may consider them
as  segments of straight lines.  We shall assume that both molecules have the same number of
base-pairs that can be visualized as points on a straight line parallel to the axis of the molecule,
one base-pair being located at the center of a corresponding molecule.  The centers of the straight
lines describing the molecules belong to the straight line perpendicular to the plane x-y parallel to
either of them. We shall denote by $\xi$ the angle between the straight lines describing the
molecules.  We assume that both molecules  are of the same helicity. The latter is determined by the
rotation of the frame of the dipole moments. Thus, we model a molecule of the DNA on a one-dimensional
lattice having at its sites either vectors of dipoles of the base pairs or scalars of the phosphate
charges. The case of total neutralization of phosphate charges was considered in paper \cite{kik}.

The dipoles are suggested to have the helix symmetry with  $\pi / 5$ rotation / bp, corresponding to
the structure of the ideal double helix of the DNA.  The distance between the centers $\kappa$, which
is fixed, is an important parameter of the model. In what follows we use the distance between adjacent
base-pairs, that is $3.4 \; \AA$, as the unit of length, and take a unit of charge for which the
dipole moment of $1 \; Debye$ equals $1$. We perform calculations in the dimensionless units generated
by these quantities.

The energy of electrostatic interaction of two molecules can be cast in the sum
\begin{equation}
    \label{U}
     \epsilon \, U = U_0 + u_{dd} + u_{dc} + u_{cd} + u_{cc}
\end{equation}
in which $\epsilon$ is the dielectric permeability of solvent and $U_0$ is the self energy of the
pair, which does not influence its conformation, the first term describes the interaction between
dipoles of the first molecule and those of the second; the second term - dipoles of the first and
phosphate charges of the second; the third - charges of the first and dipoles of the second; the
fourth - charges of the first and the second. The interactions are given by the equations
\begin{eqnarray}
    u_{dd}(\vec r, \vec r^{\; \prime}) &=& e^{-\nu \,\rho} \,
                     \left[ g(\rho) \frac{1}{\rho^3}
                             (\vec p \cdot \vec p^{\; \prime})    - 3 h(\rho)
                             \frac{ [\vec p \cdot (\vec r - \vec r^{\; \prime})]
                                    [\vec p^{\; \prime}
                                    \cdot (\vec r - \vec r^{\; \prime})]
                                  }
                                  {\rho^5}
                     \right]   \label{Udd}   \\
    u_{dc}(\vec r, \vec r^{\; \prime}) &=&  e^{-\nu \,\rho}  k(\rho)  Q
                                                       \frac{\vec p \cdot (\vec r - \vec r^{\; \prime})}{\rho}
                                                             \label{Ucd}\\
    u_{cd} (\vec r^{\; \prime}, \vec r) &=&  e^{-\nu \,\rho}   k(\rho)  Q
                                                       \frac{\vec p^{\; \prime} \cdot (\vec r - \vec r^{\; \prime})}{\rho}
                                                             \label{Udc}\\
    u_{cc}(\vec r, \vec r^{\; \prime}) &=&  e^{-\nu \,\rho}   \frac{ Q^2}{\rho}
                                                             \label{Ucc}
\end{eqnarray}
in which $\nu$ is the inverse Debye length $\nu = \lambda^{-1}$, and
$$
    \rho = |\vec r - \vec r^{\; \prime}|
$$
We shall take the screening functions $k(\rho), g(\rho), h(\rho)$ in
Schwinger's form
$$
     k = g = 1 + \nu \, \rho, \quad h = 1 + \nu \, \rho + \frac{1}{3} \, \nu^2 \rho^2
$$

The important point about the electrostatic interaction between
molecules of the DNA is  a wise choice of the screening factor.
The common practice  is to employ  the Debye-H\"uckel theory, or
its modifications that might accommodate the dipole charges, the
so-called Schwinger screening. The full treatment of this problem
requires a separate investigation. In this paper we confine
ourselves to the Debye-H\"uckel and the Schwinger theories,
\cite{schwinger}.

\section{\label{sec:group} Symmetry group for pair-interaction.}

In what follows we shall use the standard technic of group theory,  which enables us to formulate the
properties of minima for potential $U$ in a concise and clear form.

  It is worth noting that the pair potential $U$ is invariant as regards the change of
helicity: if we change the sign of angle $\xi$ between the axes of the two molecules,   at the same
time as the sign of helicity, the potential $U$ remains the same. Thus, there is the invariance rule
as regards the transformation
\begin{equation}
 \label{helicity}
        (\mbox{right helicity} , \quad \xi) \; \Leftrightarrow \; (\mbox{left helicity},
                                                               \quad - \; \xi)
\end{equation}

There are  symmetry rules for the helixes of the same kind. One may convince oneself, for example by
writing explicit expressions for the interactions given by equations (\ref{Udd})-(\ref{Ucc}) with the
help of angles $\phi_1, \phi_2$ between the dipole moments at the centers of the two molecules and
axis $z$, and the angle  $\xi$ introduced above, that the following transformations
\begin{eqnarray}
     \label{t1}
        t_1 : \;
        (\phi_1, \; \phi_2, \; \xi) & \rightarrow & (\phi_1, \; \pi - \phi_2, \; \xi + \pi)  \\
    \label{t2}
        t_2 : \;
        (\phi_1, \; \phi_2, \; \xi) & \rightarrow & ( \pi - \phi_1, \; \phi_2, \; \xi + \pi)  \\
    \label{t3}
        t_3 : \;
        (\phi_1, \; \phi_2, \; \xi) & \rightarrow & (\phi_2 + \pi, \; \phi_1 + \pi, \; \xi)
\end{eqnarray}
leave the potential $U$ invariant. The angles are defined within limits
$$
     - \pi \le \phi_1 \le \pi, \quad
     - \pi \le \phi_2 \le  \pi, \quad
     - \pi \le \xi \le \pi
$$
values  $\pm \pi$  corresponding to the same configurations of the
molecules.

The transformations given by equations (\ref{t1}-\ref{t3}) verify
the equations
$$
    t^2_1 = t^2_2 = t^2_3 = id, \quad t_2 \, t_3 = t_3 \, t_1, \quad t_1 \, t_2 = t_2 \, t_1,
$$
where $id$ is a transformation that leaves all $\phi_1, \phi_2, \xi$
invariant. Using the above equations one can easily convince oneself
that $t_1, t_2, t_3$ generate a {\it non-commutative group} of 8-th
order, ${\cal S}$. Its maximal subgroup ${\cal H}$ is a normal
subgroup of 4-th order, commutative, and generated  by the
transformations
\begin{equation}
   \label{f12}
      f_1 = t_3, \quad f_2 = t_1 \, t_2 \, t_3
\end{equation}
Elements $f_1, \; f_2$  in its turn generate subgroups ${\cal H}_1$ and ${\cal H}_2$ of ${\cal H}$,
respectively. It is worth noting that ${\cal H}_1, \;{\cal H}_2$ are of second order, both. They are
conjugate subgroups of {\cal S}, that is for an element $g$ of ${\cal S}$ we have $f_1 = g^{-1} \, f_2
\, g$, or we may state ${\cal H}_1 = g^{-1} \, {\cal H}_2 \, g$, in the notations of group theory,
which can be cast in the form of the diagram
\begin{equation}
    \label{conjug}
        {\cal H}_1 \longleftrightarrow {\cal H}_2
\end{equation}
The element
\begin{equation}
    \label{f3}
        f_3 = t_1 \, t_2
\end{equation}
generates subgroup ${\cal H}_3$ of ${\cal H}$. It is important
that ${\cal H}_3$  is a normal subgroup of ${\cal S}$, that is
$g^{-1} \,  {\cal H}_3 \, g = {\cal H}_3 $ for any element $g$ of
${\cal S}$. Thus, we have the diagram of subgroups inside the
symmetry group ${\cal S}$
\begin{equation}
    \label{groupdiagram}
    \begin{array}{lllll}
        {\cal H}_1            &          &                 &                  \\
         & \; \searrow        &          &                 &                  \\
        {\cal H}_3 & \longrightarrow & {\cal H} & \longrightarrow & {\cal S}  \\
         & \; \nearrow        &          &                 &                  \\
        {\cal H}_2            &          &                 &
    \end{array}
\end{equation}
in which the arrows signify the imbedding of subgroups.

The group of symmetries, ${\cal S}$, plays the key role in finding minima of  the potential $U$. The
following general arguments, based on the theory of groups, are quite useful in this respect. Consider
a point $\mu$ of  space ${\cal X}$ of the angles $\phi_1, \, \phi_2, \, \xi$. Suppose that $\mu$ is a
minimum of $U$. Then points
$$
    \mu^{\prime} = g \cdot \mu,
$$
called the orbit of the point $\mu$ under the action of the group
${\cal S}$, are also minima of $U$. The number of points
$\mu^{\prime}$ of  the orbit can vary. In fact, let us consider
all transformations $g$ of ${\cal S}$ that leave $\mu$ invariant,
that is $\mu^{\prime} = g \cdot \mu = \mu$. It is alleged to be
known that the transformations form a subgroup of ${\cal S}$,
called stationary subgroup ${\cal H}_{\mu}$. The stationary
subgroups, ${\cal H}_{\mu}$ and ${\cal H}_{\nu}$ , for points
$\mu$ and $\nu$ of an orbit, are conjugate, that is $ {\cal
H}_{\mu} = g^{-1}\, {\cal H}_{\nu} \, g$ for an element $g$ of
${\cal S}$. The number of different points $\mu^{\prime}$ equals
to the ratio of the orders of ${\cal S}$ and ${\cal H}_{\mu}$,
that is to 2 or 4, depending on the choice of point $\mu$. To be
specific, consider a point $\mu$ having a stationary subgroup
${\cal H}_{\mu}$ that coincides with the subgroup ${\cal H}$. The
latter is a normal subgroup of ${\cal S}$ of index 2, that is the
factor set ${\cal S} /{\cal H}$ consists of two elements. Thus,
the orbit of $\mu$ under the action of ${\cal S}$ consists of only
two points that correspond to the same value of $U$ and have the
same stationary subgroup ${\cal H}$, because the latter is a
normal subgroup of ${\cal S}$. The situation is quite different if
we take a point $\nu$ having stationary subgroup ${\cal H}_1$,
which is different from ${\cal H}_2$. The subgroups do not
coincide in ${\cal S}$, even though they are conjugate. The orbit
of ${\cal \mu}$ under the action of ${\cal S}$ indicated above
consists of four points that we may sort out as follows: two
points having the stationary subgroup ${\cal H}_1$ and two points
having ${\cal H}_2$.  This is due to the fact that for one thing
the subgroup ${\cal H}$ is commutative and therefore its elements
generate points of the orbit but with the same stationary
subgroup, that is ${\cal H}_1$, and for another there is an
element $g$ that gives points of the orbit having the stationary
subgroup ${\cal H}_2$.  In contrast, a point $\mu$ having the
stationary subgroup ${\cal H}_3$ has the orbit consisting of four
points which have the same stationary subgroup ${\cal H}_3$,
because the latter is a normal subgroup of ${\cal
S}$.\label{statsubgr}

\section{\label{sec:miimization} Minimization of potential for pair-interaction.}

It is to be noted that numerical evaluation of the minima runs across a poor convergence of standard
algorithms for minimization, because of  rather flat  surfaces of constant value for the function of
three variables, $U(\phi_1, \; \phi_2, \; \xi)$. To some extent, one may get round the difficulty by
observing that for points that remain fixed with respect to a subgroup ${\cal G}$ of ${\cal S}$, the
minimization problem is reduced to that for a smaller number of variables. This is due to the fact
that for degrees of freedom perpendicular to the set of invariant points the necessary conditions for
extremum are verified automatically and one needs only to study the conditions for  longitudinal
variables, that is to solve a smaller system of equations.

To see the point let us consider a function $f(x,y,z)$ of variables $x, y, z$ even in $x$, so that
$f(x,y,z) = f(- \, x, y,z)$. The set of invariant points is $y-z$ plane, and we may look for minima of
the function $f(x=0,y,z)$, thus we need to solve  only the two equations
$$
    \frac{\partial}{\partial y}f(x=0, y,z) = 0, \quad \frac{\partial }{\partial z}f(x=0, y,z) = 0
$$
For larger groups of symmetries the number of variables necessary for calculations can be reduced even
further, as happens for the minimization of  $U(\phi_1, \phi_2, \xi)$ we are studying. It is easy to
convince oneself that the sets of fixed points $(\phi_1, \phi_2, \xi)$, that is invariant under the
action of a subgroup of ${\cal S}$, read as follows
\begin{eqnarray}
    {\cal F}_1 &:&  (\phi_1 = \phi, \; \phi_2 = \phi + \pi, \; \xi) \label{F1}   \\
    {\cal F}_2 &:&  (\phi_1 = \phi, \; \phi_2 = - \, \phi , \; \xi) \label{F2} \\
    {\cal F}_3 &:&  (\phi_1 = \pm \, \frac{\pi}{2}, \; \phi_2 = \pm \,\frac{\pi}{2} , \; \xi)
    \label{F3}
\end{eqnarray}
in which the ${\cal F}_i$ are invariant under the action of subgroups ${\cal H}_1, \; {\cal H}_2, \;
{\cal H}_3$, respectively.
\begin{center}
 \begin{figure}[h]
     \includegraphics*[width=7cm]{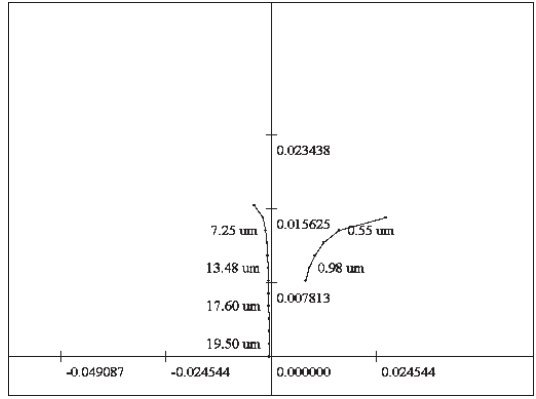}
     \caption{\label{chargeagainstenergy}
       Charge $Q$ against energy $E$ for  minima of $U$.
       Curves I, II, III correspond  to minima of the types I, II, III, respectively.
     }
 \end{figure}
\end{center}
The analysis of symmetries of  $U$ given above enables us to sort
out the minima according to the effective value of the phosphate
charge $Q$, determined by the Debye-H\"uckel screening.

\begin{center}
 \begin{figure}[h]
     \includegraphics*[width=7cm]{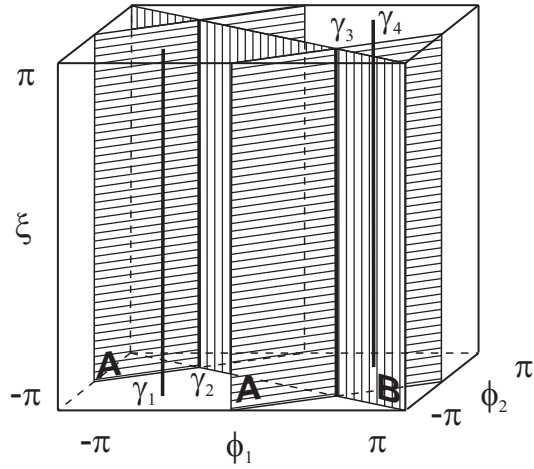}
     \caption{\label{symmetrygraph}
          Cube of the symmetries indicating sets in space $\phi_1, \phi_2, \xi$
          invariant with respect to  subgroups of ${\cal S}$.
          Main diagonal plane, $B$, corresponding to subgroup ${\cal H}_2$;
          two rectangles $A$ perpendicular to $B$ to ${\cal H}_1$;
          solid lines $\gamma_1, \; \gamma_4$, and $\gamma_2, \; \gamma_3$
          corresponding, to ${\cal H}_3$ and ${\cal H}$, respectfully.
     }
 \end{figure}
\end{center}

\begin{center}
 \begin{figure}[h]
     \includegraphics*[width=7cm]{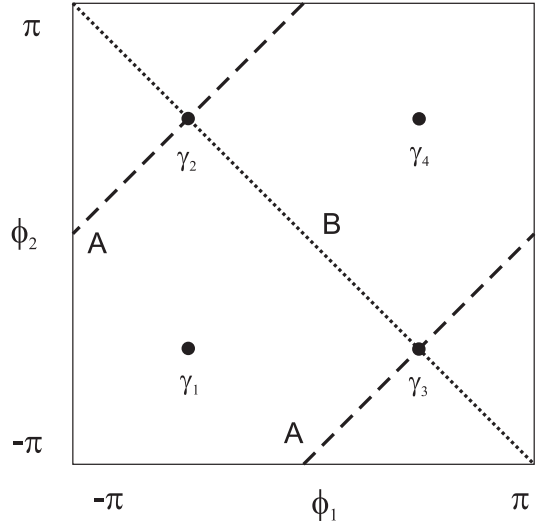}
     \caption{\label{symmetrygraph2}
          Cube of the symmetries view from the top.
          Dotted line describes invariant points ${\cal F}_2$, corresponding to
          subgroup ${\cal H}_2$; dashed line points ${\cal F}_1$ and
          subgroup ${\cal H}_1$;
          solid circles $\gamma_1$ and $\gamma_4$  to subgroup ${\cal H}_3$;
          $\gamma_2$ and $\gamma_3$  to subgroup ${\cal H}$.
     }
 \end{figure}
\end{center}

\begin{center}
 \begin{figure}[h]
     \includegraphics*[width=7cm]{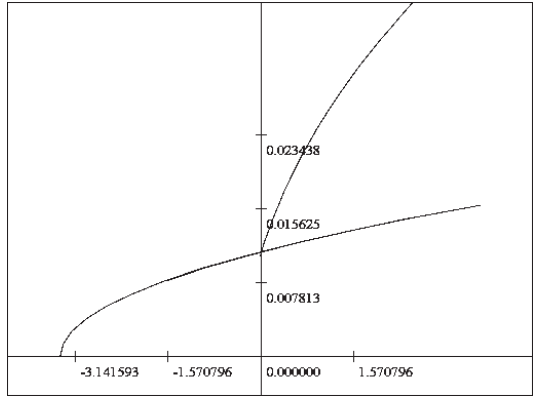}
     \caption{\label{q_E}
       Charge $Q$ against $\xi$ for minima of U.
       Curves II, III correspond to minima II, III.
    }
 \end{figure}
\end{center}

It should be noted that minima of the pair-interaction $U$ depend
on the distance between molecules $\kappa$, and the  effective
phosphate charge $Q$. The latter is the control parameter we
employ in numerical simulation. It is also useful for the
description of possible experimental results. In this paper we are
considering $\kappa$ to within $10.2 - 34 \; \AA$. Effective
charge $Q$ of phosphate groups determines the neutralization; it
varies to within $0 - .6$, in dimensionless units, $Q=0$
corresponding to the total neutralization. Charges $Q$ that
correspond to the charge inversion, have not been considered.  The
Debye length, $\lambda$, has been varied to within $7 - 35 \;
\AA$, depending on the ion strength of solution.

The numerical data and the symmetry analysis given above suggest that there should be the following
three types of the minima. \vspace{-3mm}
\begin{enumerate}
    \item Type I characterized by the molecules having a
          cross-like conformation, " snowflakes", that is $\xi$ being  close to $\pi /2$.
          It exists for $Q$ large enough.
          Its symmetry subgroup depends on the value of $Q$ and
          may take values ${\cal H}_1, \; {\cal H}_2, \; {\cal H}_3, \;{\cal H}$.
          Therefore, we may claim that there exist four sub-types of minima I:
          $\rm{I}_{{\cal H}_1}, \; I_{{\cal H}_2}, \;I_{{\cal H}_3}, \;I_{{\cal H}}$,
          each of them consisting of two subtypes which are given by specific conformations of
          the angle variables.
          \vspace{-5mm}\\
    \item Type II for which $\xi$ taking  values to within $0.1^{o}$.
          Their symmetry subgroups are ${\cal H}_1$ and ${\cal
          H}_2$, each type consisting of two sub-types.
    \item Type III for which $\xi$ is to within
          $1^{o}$, that is larger than for II. The symmetry
          subgroup is ${\cal H}_3$, and there are four constituent types of the same symmetry.
 \end{enumerate}

\section{\label{sec:concl} Liquid crystalline phases of DNA.\vspace{-7mm}}

The minima of the pair-interaction could provide a useful clue to the problem of finding possible
conformations of large ensembles of molecules, even though the pair-interaction  we have employed is a
rough approximation for studying closely packed suspensions. it is as much as to say that the
molecules in dispersions be packed in such a way that the pair-interaction should take on the lowest
possible values. Following these guidelines we feel that  the study of its minima could   allow
certain conjectures concerning liquid crystalline phases of the DNA. In particular, we may suggest
that there is a number of different phases, some of them being the cholesteric ones. In this respect
our results are in agreement:\vspace{-3mm}
          \begin{enumerate}
              \item
                   with papers  \cite{livolant} and
                   \cite{y2} in which it has been verified that the cholesteric phases are related to
                   small or almost zero effective charges of the DNA, \\
              \item
                   with the theoretical conclusions of papers \cite{lk_model} - \cite{lk3}
                   concerning the electrostatic mechanism of generating
                   cholesteric phases of the DNA,
              \item
                   with paper \cite{y2} as regards the existence of polymorphism or
                   multistability.
          \end{enumerate}
          We suggest that the description of cholesteric phases of the DNA should take into account
          the symmetries described by group ${\cal S}$ for the electrostatic interaction
          for pairs of molecules of the DNA.
          Our results are described by a sequence of the embedded
          subgroups given by diagram (\ref{groupdiagram}).

          The symmetry classification of the minima, and possible phases, enables us to see
          differences between liquid crystalline phases which at first sight appear to be physically
          identical. Thus the snowflake phase, which seems unique, if superficially estimated
          with the value of angle $\xi$, may have several different types.  The same is true
          for  "cholesteric" phase II. In contrast, cholesteric phase III is unique,
          in the sense that its symmetry subgroup is only ${\cal H}_3 $. The subgroups
          provide a concise means for describing conformations of pairs of DNA molecules
          that determine the liquid crystalline phases.

          The  physical parameter we use in our numerical work, is the effective charge
          $Q$, due to the phosphate charge of the DNA and the Debye-H\"uckel screening. It enables us,
          at least qualitatively, to take into account the experimental setting. In fact, the charge
          distribution, mimicked by $Q$, can be effected by changing the constituency of solvent.
          The numerical results  illustrated in FIG.\ref{chargeagainstenergy} are very suggestive.

          We see that for $Q$ larger than
          a certain threshold value $Q_C$ the energy of the minima corresponding to phase I,
          called the snowflake phase, is
          smaller than that corresponding to phases II and III. Therefore we may suggest that for
          solvents that provide sufficiently high value of the effective charge $Q$ there are no
          cholesteric phases {\it proper} and the snowflake phase could be present.

          It is surprising  that liquid crystals with the symmetry properties described in this paper
          have so far remained unobserved.  At least in part it might be due to the fact that the
          snowflake phase could be  misinterpreted in the X-ray experiments. It is worth noting that
          the formation of the snowflake phase is similar to that of
          the Wigner crystal, \cite{wigner},  in that it is
          due to the long-ranged Coulomb interaction of particles at low density,
          the inter-particle repulsion
          leading to the conformational organization.

          The phase diagram illustrated in
          FIG.\ref{chargeagainstenergy} describes the behavior of the effective  charge at {\it constant
          inter-helical distance}. Therefore,  special precautions
          are to be taken so as not to change the latter while  modifying the effective charge,
          with a view to see the  transition of the snowflake phase into the cholesteric one.

          It is also worthwhile to look for different cholesteric phases at small
          values of $Q$, that is less than $Q_C$.  The guideline being given by the fact
          that the size of $\xi$ differs by orders of magnitude for
          the phases II and III,  at the same time values of the effective charge $Q$ being equal.
          According to FIG.\ref{q_E} one may expect  phase transitions
          in which the cholesteric angle $\xi$ changes
          both its order of magnitude and sign.

          The formation of liquid crystalline phases of the DNA involves  the degeneracy of
          ${\cal S}$-symmetry, in accord with  the general prescription  of condensed matter theory,
          larger subgroups corresponding to more isotropic phases. For the phases under consideration
          in the present paper it could be effected owing to the increment of frustration of the
          molecules trying to satisfy both the packing and electrostatic constraints.  The symmetry
          considerations playing the key part, one may expect there could be
          phase transitions accompanied by  changes
          in symmetry described by subgroups of group ${\cal S}$.

          One may attempt to employ the conformation of charges of the  molecule  of the DNA that
          relies on the pyrophosphate groups rather than on the phosphate ones as usual. By now
          there has been a considerable progress in studying  molecules of the DNA with
          substituted pyrophosphate internucleotide groups,\cite{kuz1}.
          The samples studied in paper \cite{kuz1} had the double charges,
          $- 2$ electron charge, due to the pyrophosphate groups inserted periodically, usually one group
          for every ten base-pairs of the DNA. One can expect that the study of dispersions
          of the DNA based on  molecules carrying pyrophosphate internucleotide  groups, could make
          for understanding the part played by electrostatic forces in generating liquid crystalline
          phases of the DNA. The design of novel DNA probes, \cite{kuz2}, provides new opportunities
          in this respect.

          The authors are thankful to S.Ya.Ishenko, S.A.Kuznetsova, and Yu.M.Yevdokimov for discussions.
          The work has been partially financed by the MSU Interdisciplinary Project for Studying
          Macromolecules.

\end{document}